\documentclass[letterpaper]{article}
\usepackage{aaai2027}
\nocopyright
\usepackage[hyphens]{url}
\usepackage{graphicx}
\usepackage{natbib}
\usepackage{caption}
\usepackage{booktabs}
\usepackage{amsmath}
\usepackage{xcolor}

\title{Compact Documentation for Coding Agents: A Benchmark, an Optimizer, and Why It Does Not Transfer}
\author{
    Md Shohel Arman\textsuperscript{\rm 1},
    Igor Molybog\textsuperscript{\rm 2}
}
\affiliations{
    \textsuperscript{\rm 1}Daffodil International University, Dhaka, Bangladesh\\
    \textsuperscript{\rm 2}HawAII, University of Hawai`i at M\=anoa, Honolulu, HI, USA\\
    arman.swe@diu.edu.bd, molybog@hawaii.edu
}
\begin{document}
\maketitle
\begin{abstract}
We investigate whether natural-language documentation helps coding agents resolve software issues, and we build the tools to construct and evaluate it. We introduce a roundtrip benchmark that scores code descriptions by whether code regenerated from them passes the original tests, and show that completeness, not length, drives a description's fidelity. Using the benchmark as an optimization signal, we discover a description-writing prompt that reaches full fidelity and generalizes to unseen files. We then test the hypothesis that motivated the work: that better documentation helps an agent resolve real repository issues. Across two model families and ten repositories, and against a positive control confirming that our evaluation can detect a genuine improvement, we find that it does not. When the source is present, neither static compact documentation nor retrieved context beats the issue alone. We report this negative result together with the benchmark and the optimizer, and we characterize the boundary at which documentation helps.
\end{abstract}
\renewcommand{\thefootnote}{}\footnotetext{Preprint. Code and data: \url{https://github.com/haw-ai-i/roundtrip}}\setcounter{footnote}{0}
\section{Introduction}

For large enough repositories, loading the entire codebase into an agent's context is prohibitively expensive or impossible. Coding agents must therefore edit files they have not read in full, relying instead on high-level descriptions in their context to supply the intent and contracts that the raw code leaves implicit. Agents operate this way on real repositories \citep{sweagent2024, autocoderover2024}, but an incomplete or incorrect description acts as a faulty map. This motivates a natural hypothesis: high-fidelity descriptions, standing in for source the agent cannot load, should help it resolve real issues. We set out to build documentation good enough to serve this role, and to test whether it does. Our focus is on evaluating and optimizing the descriptions themselves, independent of the agent's internal machinery, so we first investigate what a good description contains and how to generate it automatically, and then ask whether better descriptions actually help downstream.

We approach the problem in four steps. First, we establish when documentation helps an agent and identify the two properties governing its value: completeness and compactness. Second, we develop a roundtrip benchmark that measures these properties by regenerating code from descriptions and running the original tests, validating the benchmark on real repositories while controlling for memorization. Third, we use the benchmark as an optimization signal and discover a description-writing prompt that reaches full regeneration fidelity and generalizes to unseen files. Fourth, and this is the question that motivated the work, we test at scale whether these descriptions help an agent resolve real repository issues.

Our central claim about description quality holds: completeness, not length, is the primary driver of a description's fidelity, and the optimizer confirms it by converging to prompts that are both complete and compact. The downstream claim does not hold. When the source file is withheld, an optimized description substitutes for it well enough to lift issue resolution from a mean test-pass fraction of 0.08 to 0.71. But when the source is present, which is how coding agents normally operate, the benefit disappears. Across two model families and ten repositories, and against a positive control confirming that our evaluation can detect a genuine improvement when one exists, neither a static compact description nor retrieved past-task context improves issue resolution over giving the agent the issue alone; longer documentation can make matters slightly worse (Table~\ref{tab:present}). We report this negative result honestly, alongside the benchmark and optimizer that produced it, and we locate the boundary separating the two regimes.

We make three contributions. (1) A roundtrip benchmark that scores a describer, the system producing a natural-language description, by whether code regenerated from its description alone passes the original tests, together with a compression measure and controls for memorization. (2) A controlled account of when documentation helps an agent, separating the effect of completeness from that of length, and a self-improvement loop that optimizes the describer against the benchmark, reaches full fidelity, and generalizes to held-out files. (3) A large-scale, multi-model test of the downstream hypothesis that better documentation helps issue resolution, which, with a validating positive control, returns a robust null in the source-present setting and locates the boundary at which documentation stops helping.

\section{Background}

\paragraph{Roundtrip and self-consistency evaluation.} Our benchmark builds most directly on round-trip correctness, which evaluates a code model by describing code in natural language and regenerating code from that description, using existing unit tests as the equivalence oracle \citep{rtc2024}. IdentityChain shows that self-consistency and accuracy are distinct, and that many code models fail to preserve their own behavior across the roundtrip \citep{identitychain2024}. More recent work frames roundtrip consistency through invertibility, finding that models can succeed at the forward and backward directions individually while failing their composition \citep{rtce2026}. We differ in turning the roundtrip from a diagnostic into an optimization target: we not only measure whether a description regenerates its code, but use that signal to improve the description writer.

\paragraph{Execution-based code benchmarks.} Execution against real test suites is the standard for evaluating code models, from function-level benchmarks such as HumanEval \citep{humaneval2021} and MBPP \citep{mbpp2021} to repository-scale evaluation. SWE-bench evaluates whether models resolve real repository issues, using fail-to-pass and pass-to-pass tests in isolated containers \citep{swebench2024}; our fixtures adopt the same single-file, test-as-oracle discipline. A recent line of work examines the role of context in this setting: SWE-ContextBench measures whether agents reuse retrieved past-task experience across related issues \citep{swecontextbench2026}, and ContextBench annotates gold contexts to evaluate how agents retrieve and use code context during resolution \citep{contextbench2026}. Our downstream experiments use SWE-ContextBench directly, and ask a complementary question: whether static, issue-independent documentation of a file helps once that file's source is already available. Repository-level benchmarks such as MRG-Bench \citep{mrgbench2025} and CoderEval \citep{codereval2024} show that non-standalone, context-dependent code is substantially harder than isolated functions, and benchmarks that generate whole libraries or services from natural-language documentation report low functional pass rates even for strong models \citep{nl2repobench, repogenesis}.

\paragraph{Code documentation and compression.} A long line of work generates natural-language documentation from code \citep{docgpt3, chatgptsumm, doccompare}. Closest to our compactness result is work on documentation compression: that shorter docstrings can retain code-generation performance \citep{lessismore2024}, that a function signature may carry most of the needed information \citep{signatureenough2024}, that docstring reformulation can improve generation \citep{docreform2024}, and that prompts can be compressed with little loss \citep{llmlingua2023}. Our experiments give a controlled account of this: uplift tracks completeness, while verbosity beyond completeness adds nothing.

\paragraph{Prompt and harness optimization.} Our optimization loop follows the tradition of automatic prompt optimization: generating and selecting candidate instructions \citep{ape2023}, refining them with prior scores in context \citep{opro2023}, compiling declarative pipelines into optimized prompts \citep{dspy2023}, evolving prompt populations \citep{promptbreeder2023}, and back-propagating textual feedback \citep{textgrad2024}. We apply this to documentation quality specifically, scored by regeneration fidelity, following the harness-search framing of \citet{autoagent} and \citet{metaharness}. Table~\ref{tab:related} places our benchmark next to the closest prior work.

\begin{table*}[t]
\centering
\begin{tabular}{@{}lllll@{}}
\toprule
Work & Direction & Granularity & Oracle & Optimizes writer? \\
\midrule
RTC \citep{rtc2024}            & code$\to$NL$\to$code & function & unit tests        & no \\
IdentityChain \citep{identitychain2024} & code$\leftrightarrow$NL & function & test-output match & no \\
RTCE \citep{rtce2026}          & invertible round trip & snippet  & exact match       & no \\
SWE-bench \citep{swebench2024} & NL$\to$code          & repo     & fail/pass tests   & no \\
MRG-Bench \citep{mrgbench2025} & context$\to$code     & repo     & runnable tests    & no \\
CoderEval \citep{codereval2024}& NL$\to$code          & function & unit tests        & no \\
This work            & code$\to$NL$\to$code & file     & unit tests        & yes \\
\bottomrule
\end{tabular}
\caption{Positioning against related benchmarks. Prior roundtrip and execution-based benchmarks measure a model or a description; ours additionally uses the roundtrip score to optimize the system that writes the description.}
\label{tab:related}
\end{table*}

\section{When Documentation Helps an Agent}
\label{sec:motivation}

\subsection{Experimental Setup}

All generation runs at temperature zero unless otherwise noted, and for each condition we report the number of independent attempts that pass the associated oracle. Model names are given with each experiment. Appendix C of the supplementary material gives the full set of prompts used across these experiments.

\subsection{Documentation Helps Only Beyond the Code}

Before optimizing descriptions, we establish when a description helps an agent at all. We find that documentation helps only when it carries intent or a contract the code does not expose; a description that restates readable code adds nothing. Flash-lite performs the task on a codebase, once with the code alone and once with the code plus a description written by Pro, and we compare success. Flash-lite is used here so that success depends on the information the description supplies rather than on raw capability.

The description helps precisely when it carries information the code does not locally expose. We hand-write three small codebases in which the intended behavior is a \emph{contract} that the code does not enforce, and we write the description for each by hand so that it states the contract in words while the code does not. The task asks the agent to add a method that honors the contract. A concrete example is an access-control policy engine. Its \texttt{is\_allowed} method simply scans the rules in the order they were supplied and returns the first match; the intended contract, that rules are evaluated by priority and that at equal priority a deny overrides an allow, is stated only in the description and is nowhere visible in the code's behavior. The task asks the agent to add a \texttt{resolve} method that applies this contract regardless of input order. Because the contract cannot be inferred from the code, the agent must obtain it from the description. Adding the description flips the weak agent from consistent failure to consistent success, from zero of three attempts to three of three on each of the three codebases. When the description instead merely restates code the agent can already read, including a real 423-line source file, it adds nothing: the agent reads the logic directly and succeeds with or without the description. Documentation thus helps to the degree it supplies intent or contract beyond what the code shows. This also implies a tension we return to later: a maximally faithful description of visible, correct code is by construction redundant with that code.

\subsection{What Determines a Description's Value}

Given that a description can help, two properties determine how much. We find that completeness decides the agent's success, while length has little effect once completeness is achieved. We isolate each with a comprehension task whose implementation is trivial, so that success depends only on whether the description conveyed the needed facts rather than on the agent's coding ability. Here a \emph{fact} is a single required key--value pair: a configuration setting whose exact name and literal value the task must recover (for example, a default such as \texttt{max\_retries = 5}). The task asks the agent to return five such documented values that appear only in the description, so a correct answer requires only a lookup and the sole variable is whether the description carried the values.

For completeness, we build descriptions that carry a graded fraction of the five facts: we include exactly one, three, or five of the required key--value pairs and omit the rest, holding the surrounding prose fixed. Uplift tracks completeness exactly: descriptions carrying one, three, or five of five facts yield one, three, and five correct, and a description carrying none yields none. The agent recovers precisely the facts the description states, and nothing more.

Completeness clearly matters. Length is a natural second hypothesis: perhaps a long description, even a complete one, hurts a weaker model, since models are known to use information less reliably as the surrounding context grows \citep{lostmiddle2023}. We tested this directly. In a first comparison, a compact 38-word description and a 664-word one carrying the same five facts, about seventeen times longer, produced equal agent uplift (Table~\ref{tab:stage2}). We then made the test harder, using 28 configuration settings, of which ten are checked, with deliberately \emph{confusable} names, close variants such as \texttt{max\_retries}, \texttt{max\_retry\_delay}, and \texttt{connection\_max\_retries} that a model could easily conflate, and we embedded the needed keys in long filler prose, and ran the comparison across several models. We do not find that length itself hurts: a complete description performs the same whether short or long. The supplementary material reports the full set of these experiments, including one case where an apparent length effect turned out to be a missing-information artifact. We therefore treat length as a second-order factor, well below completeness. What decides an agent's success is completeness. This is closely related to what the roundtrip benchmark measures with tests: a test fails exactly when the description omits a behavior the test checks, so passing more tests reflects a more complete description, though tests probe behavior rather than enumerating discrete facts.

\begin{table}[t]
\centering
\begin{tabular}{@{}lcc@{}}
\toprule
Condition & No desc. & With desc. \\
\midrule
Contract only in description & 0/3 & 3/3 \\
Contract visible in code     & 3/3 & 3/3 \\
1 of 5 facts present         & 0/5 & 1/5 \\
3 of 5 facts present         & 0/5 & 3/5 \\
5 of 5 facts present         & 0/5 & 5/5 \\
Complete, compact (38w)      & 0/5 & 5/5 \\
Complete, verbose (664w)     & 0/5 & 5/5 \\
\bottomrule
\end{tabular}
\caption{When and why documentation helps a weak agent, as pass counts without versus with the description. Documentation helps only when it carries a contract the code does not expose; success rises with the number of facts the description includes; and a compact and a verbose description carrying the same facts give the same result.}
\label{tab:stage2}
\end{table}

\section{The Roundtrip Benchmark}
\label{sec:benchmark}

\paragraph{Data.} Our fixtures are drawn from SWE-bench Verified \citep{swebench2024}, the human-validated subset on which foundation models report their scores and which is therefore the natural decontamination target. We apply an explicit filter rather than a hand selection: an instance qualifies if its gold patch modifies exactly one non-test source file and its test patch names the oracle tests. This filter admits 429 of the 2294 Verified instances; restricting to the sympy project and to instances whose era supports a current Python toolchain, we build an environment for each of the most recent candidates, apply the instance's gold and test patches, and keep every instance whose original code passes its own oracle by construction. Environments whose baseline fails, whose era predates the toolchain, or whose test patch spans several files are excluded and logged with the reason. Eleven fixtures result, each a single module together with the instance's own unit tests, which serve as the oracle; every fixture is checked before use by scoring its own gold-patched source through the oracle, which must pass in full. The controlled boundary experiments additionally use three hand-constructed codebases, released with the benchmark, whose behavioral contract is fixed by design.

\paragraph{Contamination controls.}
Because a large model may have seen public source during training, a high fidelity score could in principle reflect memorization of the original file rather than genuine regeneration from the description. Drawing the fixtures from SWE-bench Verified addresses this at the source: Verified is the subset foundation-model providers report against and is the standard decontamination target for their training corpora. We additionally monitor the regenerate stage for verbatim recitation of the original source; on the benchmark fixtures we observed none, and the regenerated files differ from the originals while still being scored against the tests. On widely mirrored files outside the benchmark the model can instead refuse generation outright when it recognizes the source, which blocks the roundtrip rather than inflating it; such files cannot serve as fixtures and the selection excludes them.

\paragraph{Pipeline.}
We evaluate an agent harness using a three-stage pipeline. In the describe stage, the agent is tasked to generate a natural-language description of a source file. The original unit tests are excluded from the available context. In the regenerate stage, a separate model produces a new implementation using only the description and an empty project scaffold, without access to the original source code. The scaffold is the surrounding package structure with the target file removed, together with a contract file that names the public symbols the regenerated module must expose, so that other modules can import it. In the evaluate stage, the regenerated implementation is scored against the file's original test suite. Generation stages operate at temperature zero; the hosted API is nonetheless not bit-deterministic, so benchmark scores are reported as means over repeated runs.

\paragraph{Compression.}
A description that reproduces the file verbatim would score high on fidelity but defeat the purpose. We therefore also measure how compact a description is, relative to the code it describes:
\begin{equation}
\mathrm{density}(d) = \frac{\text{tokens in } d}{\text{tokens in the code described}},
\end{equation}
where both counts use the same subword tokenizer, giving a ratio that is comparable across files regardless of formatting. Density is defined per file and reported for the per-file regeneration protocol used throughout this paper, in which each file is described and regenerated on its own against a shared contract. An alternative protocol regenerates a whole package from a single prompt; because one description then covers several files, its effective density is not directly comparable to the per-file figure, so we keep the two protocols separate.

\section{Benchmark Results}

Our fixtures are drawn from SWE-bench Verified (Section~\ref{sec:benchmark}); every instance that passed the selection filter and environment verification is reported, eleven fixtures in total. Both the describe and the regenerate stage run on Flash (\textsc{gemini-3.5-flash}) at temperature zero, using the base prompts given verbatim in the supplementary material; Flash is the mid-sized model of the family, the tier at which fidelity separates descriptions. Because the hosted API is not bit-deterministic even at temperature zero, we run each fixture three times and report the mean and standard deviation (Table~\ref{tab:fixtures}). The fixtures span the full range: one regenerates perfectly on every run, several earn stable partial credit, and four fail outright on every run, so the benchmark is far from saturated and fidelity separates good descriptions from weak ones.

\begin{table}[t]
\centering
\begin{tabular}{@{}lrrr@{}}
\toprule
Fixture & Lines & Density & Fidelity \\
\midrule
contains        & 48   & 1.57 & 1.00 $\pm$ 0.00 \\
point           & 1376 & 0.56 & 0.67 $\pm$ 0.09 \\
unitsystem      & 205  & 1.22 & 0.63 $\pm$ 0.19 \\
unitsystem\_v2  & 205  & 1.30 & 0.61 $\pm$ 0.54 \\
homomorphisms   & 549  & 0.94 & 0.33 $\pm$ 0.00 \\
ast             & 1869 & 0.39 & 0.18 $\pm$ 0.32 \\
lambdify        & 1402 & 0.27 & 0.11 $\pm$ 0.19 \\
symbol          & 925  & 0.41 & 0.00 $\pm$ 0.00 \\
pycode          & 642  & 0.43 & 0.00 $\pm$ 0.00 \\
gamma\_matrices & 716  & 0.29 & 0.00 $\pm$ 0.00 \\
rings           & 2470 & 0.22 & 0.00 $\pm$ 0.00 \\
\midrule
Mean $\pm$ SD & 946 $\pm$ 760 & 0.69 $\pm$ 0.48 & 0.32 $\pm$ 0.35 \\
\bottomrule
\end{tabular}
\caption{Per-fixture roundtrip results on the eleven SWE-bench Verified fixtures. Fidelity is the fraction of original tests passed by code regenerated from the description alone, reported as mean $\pm$ standard deviation over three independent runs; density is tokens in the description divided by tokens in the code, measured on the same runs. The bottom row aggregates across fixtures.}
\label{tab:fixtures}
\end{table}

When a description fails to regenerate its file, the failure falls into a small number of recurring modes (Table~\ref{tab:failures}). A description may omit an exact output string the tests check, drop an import the code relies on, or fail to restate an invariant such as immutability that the original enforced. In a distinct mode, the regenerated file fails to parse or to import at all: on the larger fixtures the model emits a file with a syntax error, a wrong internal signature, or an import of a name that does not exist, and because these are modules the rest of the package imports during test collection, a single such defect prevents any test from running. This mode dominates the zero-fidelity fixtures: for core modules the roundtrip is all-or-nothing at import time, while leaf modules earn partial credit test by test. On our decontaminated Verified fixtures we did not observe the recitation mode, in which a model declines to reproduce source it recognizes, though it appears on widely mirrored files outside the benchmark. These modes recur across the failing fixtures rather than being one-offs, and their recurrence is what makes them worth targeting; the same modes directly inform the optimization in the next section.

\begin{table}[t]
\centering
\begin{tabular}{@{}ll@{}}
\toprule
Failure mode & Example \\
\midrule
Lost output literal & omits an exact string the tests check \\
Lost imports & drops a needed import \\
Lost constraint & an invariant is not restated \\
Recitation blocked & declines to reproduce known source \\
\bottomrule
\end{tabular}
\caption{Failure modes when a description fails to regenerate its file.}
\label{tab:failures}
\end{table}

\begin{figure}[t]
\centering
\includegraphics[width=\columnwidth]{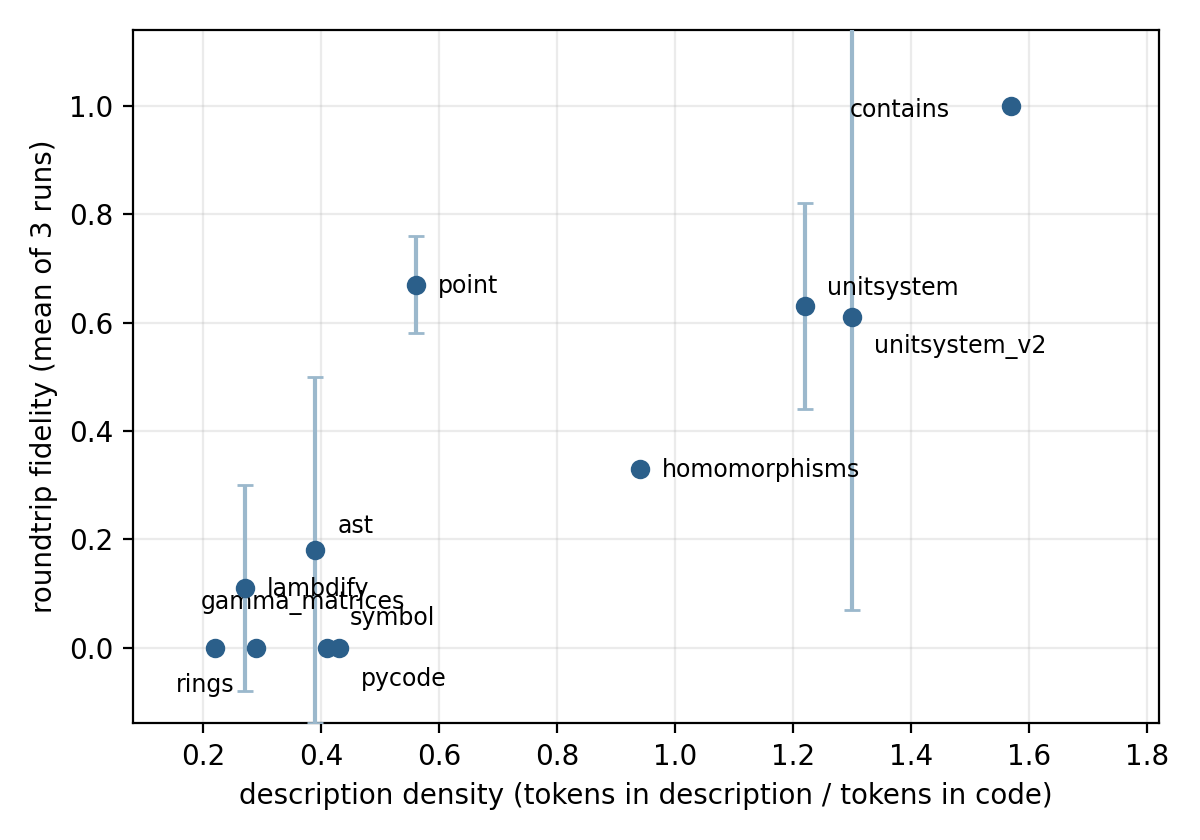}
\caption{Roundtrip fidelity (mean $\pm$ SD over three runs) against description density (tokens in the description over tokens in the code) across the eleven Verified fixtures.}
\label{fig:compression}
\end{figure}

Figure~\ref{fig:compression} plots fidelity against description density. Files whose descriptions carry more tokens per token of code regenerate more faithfully: across the eleven fixtures the Pearson correlation between density and mean fidelity is 0.88. The four zero-fidelity fixtures all have density below 0.45, while every fixture at density 0.9 or above reaches fidelity 0.33 or higher. The reading is direct: when the describe stage compresses a large file too aggressively, the description loses the internal details regeneration needs, and fidelity collapses. Density is therefore a meaningful axis of description quality alongside completeness.

\subsection{Comparison to an External Documentation Tool}

The comparisons in this subsection were run on the benchmark's original SWE-bench fixture set. The benchmark also lets us ask how descriptions written for a different purpose fare on regeneration. We compare against OpenWiki \citep{openwiki2026}, a recent open-source tool that generates documentation for a codebase specifically for coding agents to consume. OpenWiki is a natural baseline: it targets the same setting we do, an agent working from documentation rather than from the code itself, but it is built to produce browsable repository documentation rather than a specification for reconstruction.

We ran OpenWiki on the three fixtures of the original set that span its fidelity range, one high, one middle, and one low, since generating OpenWiki documentation requires a full run of the external tool with its strongest model per repository and the comparison needs the range rather than the count, and fed the documentation it produced through the same regenerate and evaluate stages we use for our own descriptions, so the only thing that changes is the source of the description. To keep the comparison on the same model family, we pointed OpenWiki at the same provider our benchmark uses; the supplementary material gives the full setup. Table~\ref{tab:openwiki} shows the result: on all three fixtures the documentation OpenWiki generated regenerates the code less faithfully than our descriptions do; the supplementary material reproduces one of our descriptions in full for comparison. The gap has a consistent cause. OpenWiki writes clear, readable documentation of each module's public interface, with usage examples and API summaries, but it leaves out internal details the tests depend on. On \texttt{saferepr} it omitted a private helper that the tests import directly, and on \texttt{unitsystem} it omitted a class attribute the code relies on, so in both cases regeneration failed before any test could run. This is the same lesson the earlier experiments gave from the other direction: what a description needs for regeneration is completeness, down to the internal details that reference documentation usually leaves implicit. A tool that produces browsable reference documentation and a describe stage tuned for reconstruction are optimized for different objectives.

\begin{table}[t]
\centering
\begin{tabular}{@{}lcc@{}}
\toprule
Fixture & OpenWiki & Our description \\
\midrule
saferepr    & 0.00 & 0.73 \\
contains    & 0.67 & 1.00 \\
unitsystem  & 0.00 & 0.85 \\
\bottomrule
\end{tabular}
\caption{Roundtrip fidelity of documentation produced by an external tool (OpenWiki) against our own descriptions, scored through the identical regenerate-and-evaluate pipeline.}
\label{tab:openwiki}
\end{table}

The same gap appears when the documentation is used to implement a code change rather than to regenerate a file. The task is the deny-override task of Section~\ref{sec:motivation}: the agent must add a \texttt{resolve} method that evaluates access-control rules by priority, with a deny overriding an allow at equal priority, a contract that is absent from the code's behavior and present only in the description. We gave Flash-lite OpenWiki's documentation in place of ours. Table~\ref{tab:openwiki-agent} shows the result across the three documentation conditions. The reason is visible in what OpenWiki wrote: it correctly reported that the engine processes rules in input order and does not use priority, which is the observed behavior but the opposite of the contract the task requires. A tool that documents what the code does cannot supply a contract the code does not exhibit, which is exactly the information an agent needs here.

\begin{table}[t]
\centering
\begin{tabular}{@{}lc@{}}
\toprule
Documentation given to Flash-lite & Passes \\
\midrule
None                & 0/4 \\
OpenWiki            & 0/4 \\
Our description     & 4/4 \\
\bottomrule
\end{tabular}
\caption{The deny-override task under three documentation conditions. Each row reports how many of four independent attempts by Flash-lite implement a \texttt{resolve} method that satisfies the priority and deny-override contract. OpenWiki's documentation, which reports the code's observed order-dependent behavior, leaves the agent exactly where no documentation leaves it.}
\label{tab:openwiki-agent}
\end{table}

\section{Optimizing the Description Writer}
\label{sec:optimization}

\subsection{Method}

\begin{figure}[t]
\centering
\includegraphics[width=\columnwidth]{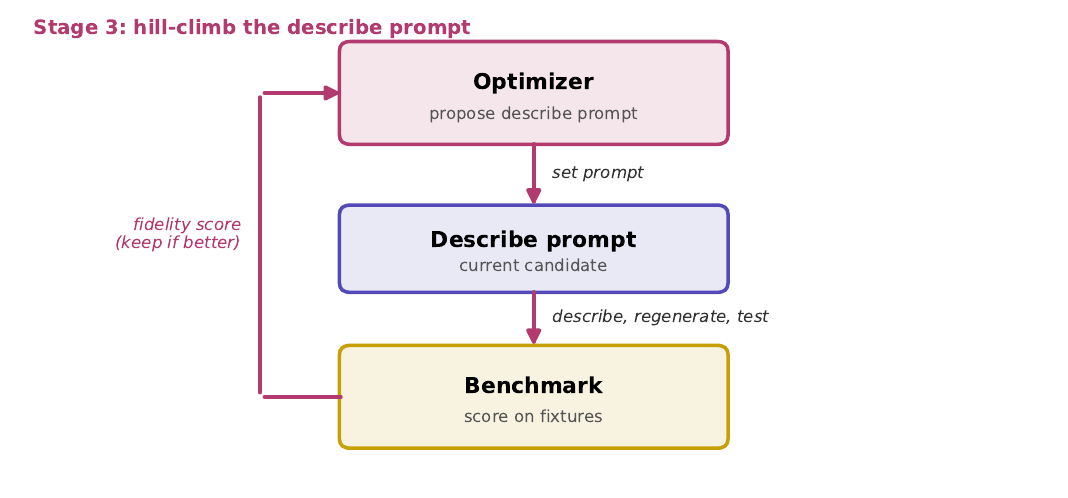}
\caption{The prompt optimization loop. The optimizer proposes a describe prompt; the benchmark scores the descriptions it produces; the prompt is kept if the score improves. Optimization uses a training set of fixtures, and the discovered prompt is then evaluated on a held-out set never seen during optimization.}
\label{fig:stage3loop}
\end{figure}

Having established that completeness is what carries a description's value, we ask whether a system can be made to write them. We treat the describe-stage prompt as a harness and optimize it against the benchmark, in the style of automated harness search. An outer loop proposes a new describe prompt, scores the descriptions it produces on a set of fixtures, and keeps the prompt if it improves. The objective rewards mean fidelity with a small penalty on description length. For a describe prompt $\pi$ evaluated over a set of fixtures $F$,
\begin{equation}
J(\pi) = \frac{1}{|F|}\sum_{f \in F} \mathrm{fidelity}(\pi, f)
\; - \; \lambda \, \frac{\bar{w}(\pi)}{300},
\end{equation}
where $\bar{w}(\pi)$ is the mean description length in words produced by $\pi$, the constant $300$ normalizes that length to a scale comparable to fidelity, and $\lambda = 0.1$. The penalty is small enough that fidelity dominates, so the loop is pushed toward descriptions that are both complete and compact. The proposer is Pro (\textsc{gemini-2.5-pro}), sampled at temperature $0.4$ so that successive proposals differ, while the describe and regenerate stages scored inside the loop stay at temperature zero and so give each candidate prompt a deterministic score. The proposer is given the current prompt together with its measured fidelity and length, and asked to rewrite it to score higher; its prompt is given verbatim in the supplementary material. To test that improvements generalize rather than overfit to the fixtures being optimized, we split our fixtures into a training set of three fixtures that the loop optimizes on and a held-out set of two fixtures it never sees during optimization. The held-out set is small, and we report the generalization as a proof of concept rather than a large-sample result. Figure~\ref{fig:stage3loop} shows the loop.

\subsection{Results}

\begin{figure}[t]
\centering
\includegraphics[width=\columnwidth]{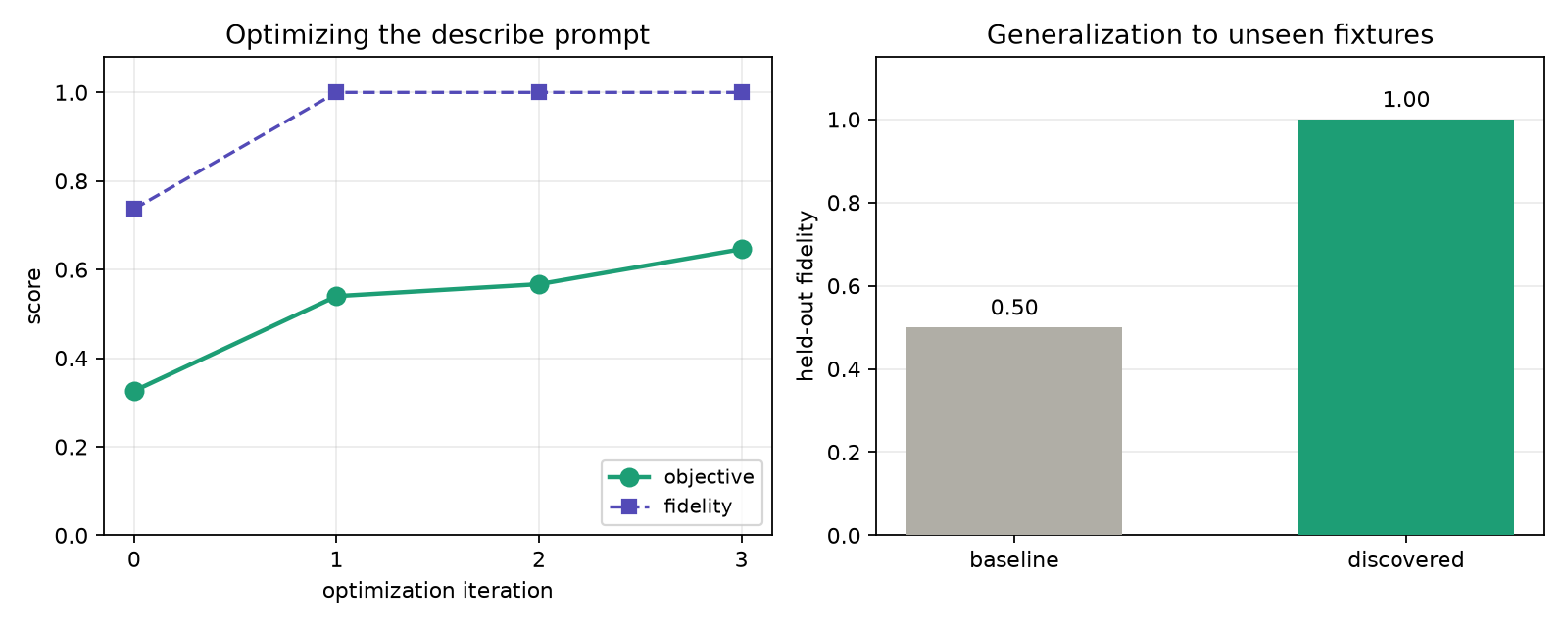}
\caption{Optimizing the describe prompt. Left: the objective and mean fidelity climb across iterations; the loop reaches fidelity 1.0, then reduces length. Right: the discovered prompt raises held-out fidelity from 0.5 to 1.0 on fixtures it never optimized on.}
\label{fig:stage3}
\end{figure}

The loop improves the prompt (Figure~\ref{fig:stage3}). Starting from a hand-written baseline, the objective climbs at every iteration. The loop first raises fidelity to 1.0 across the training set, then holds fidelity at 1.0 while cutting description length, exactly the complete-then-compact behavior the thesis predicts: once descriptions are complete, the only way to improve the objective is to make them shorter. On held-out fixtures the loop never optimized on, the discovered prompt raises mean fidelity from 0.5 to 1.0 while using fewer words than the baseline, so the improvement generalizes to unseen files. The result is reliable rather than a lucky trajectory: three independent runs with different proposer temperatures all reach full training fidelity and preserve it out of sample.

The discovered prompt is interpretable, and this is the strongest evidence that the optimization captures real signal. Where the baseline asked in general terms for inputs, outputs, and edge cases, the discovered prompt added specific instructions that map directly onto the failure modes of Table~\ref{tab:failures}: list all imports precisely, define module-level constants with their exact literal values, and state exact signatures, defaults, and the conditions for every exception and return path. The full discovered prompt and the base prompts it was optimized from are given in the supplementary material. The optimizer had no access to our hand analysis of failure modes, so the agreement between the failures we catalogued and the fixes it discovered is independent evidence that the improvements are genuine.

\section{Transfer to Issue Resolution}
\label{sec:resolution}

The optimization of Section~\ref{sec:optimization} improves the describe prompt against the roundtrip objective. The question that motivated this work is whether those gains transfer to the task the documentation is meant to serve: resolving a real repository issue, in the style of SWE-bench \citep{swebench2024}. We separate two regimes. When the source file is withheld, documentation is the agent's only view of the code, and here the optimized descriptions produce a large, real improvement. When the source is present, which is the normal operating condition of a coding agent, we test at scale across models and repositories and find no improvement at all. In every experiment the descriptions are generated from the pre-fix file, so a fix cannot leak through the documentation.

\subsection{Source withheld: documentation substitutes for code}

We first test on the eleven Verified fixtures with the source file removed. The agent (Flash) receives the issue text and, by condition, no documentation, a description from the baseline describe prompt, or a description from the optimized prompt, and must produce the complete file; the instance's own tests score the attempt, run three times per fixture. Table~\ref{tab:transfer} reports the mean fraction of tests passed. With the issue alone the agent reaches 0.08; baseline descriptions lift this to 0.21; optimized descriptions reach 0.71 and are the best condition on ten of eleven fixtures. The optimized descriptions are also far more stable across attempts. The two fixtures where the optimized condition fails, rings and lambdify, are exactly those whose roundtrip fidelity is near zero because the describe stage over-compressed them, so resolution without source tracks roundtrip fidelity. A prompt learned purely against the roundtrip objective produces descriptions that carry a module well enough for an agent to resolve real issues in it without ever seeing the code.

\begin{table}[t]
\centering
\begin{tabular}{@{}lccc@{}}
\toprule
Fixture & Issue & Baseline & Optimized \\
\midrule
contains        & 0.33 & 0.56 & 0.83 \\
point           & 0.00 & 0.00 & 0.77 \\
unitsystem      & 0.00 & 1.00 & 0.99 \\
unitsystem\_v2  & 0.00 & 0.00 & 0.97 \\
homomorphisms   & 0.00 & 0.00 & 0.22 \\
ast             & 0.18 & 0.30 & 0.88 \\
lambdify        & 0.00 & 0.00 & 0.10 \\
symbol          & 0.00 & 0.46 & 1.00 \\
pycode          & 0.00 & 0.00 & 0.95 \\
gamma\_matrices & 0.00 & 0.00 & 1.00 \\
rings           & 0.33 & 0.00 & 0.09 \\
\midrule
Mean & 0.08 & 0.21 & 0.71 \\
\bottomrule
\end{tabular}
\caption{Issue resolution with the source file withheld. Mean fraction of the instance's tests passed over three attempts. Issue gives the agent the issue text alone; Baseline and Optimized additionally supply a description of the pre-fix file from the baseline and optimized describe prompts.}
\label{tab:transfer}
\end{table}

\subsection{Source present: documentation stops helping}

The result reverses once the agent can read the code. We test this in the standard issue-resolution setting, where the pre-fix repository is in the working directory and the agent edits in place; an attempt is resolved only if the instance's fail-to-pass and pass-to-pass tests all pass. We ran this progression deliberately, moving from a weak local model to a capable hosted one, from our own fixtures to an external benchmark, and from a small set to its full extent, so that a null could not be dismissed as an artifact of any single choice. Table~\ref{tab:present} collects the outcomes.

\begin{table*}[t]
\centering
\small
\begin{tabular}{@{}llccccc@{}}
\toprule
Setting & Model & Paired & Issue only & +Compact & +Full & +Context \\
\midrule
Kind-B multi-file   & Qwen 3.6         & 43 & 14 & --  & 11 & --  \\
Kind-B django       & Gemini 3.1 Pro   & 15 & 9  & 8   & 4  & --  \\
SCB django, control & Gemini 3.1 Pro   & 29 & 14 & --  & -- & 15  \\
SCB django          & Gemini 3.1 Pro   & 32 & 15 & 16  & -- & --  \\
SCB Lite            & Gemini 3.8 Flash & 58 & 33 & 29  & -- & 30  \\
\bottomrule
\end{tabular}
\caption{Issue resolution with the source present, across models, fixture sets, and documentation types. Each cell is the number of resolved tasks out of the Paired total, the tasks that produced a scorable edit under every condition compared in that row; a dash means the condition was not run in that setting. +Compact is the roughly eighty-word static description of the target file, +Full the full-length description from the optimized prompt, and +Context the benchmark's retrieved past-task experience. For Kind-B multi-file, 43 of 50 attempted fixtures paired. The SCB django control and the SCB django compact comparison come from separate runs with different attrition, so their issue-only counts (14 and 15) are taken over different task sets. No documentation condition improves on issue-only by more than a single task, and the full-length description is the worst condition wherever it was run.}
\label{tab:present}
\end{table*}

\paragraph{A weak model on our own fixtures.} On the multi-file Kind-B set with a local Qwen 3.6 agent, adding the optimized description does not help. Of 50 fixtures attempted, 43 produced a scorable edit under both conditions; the remaining seven are excluded for an unrunnable oracle or no edit within budget. On the 43 paired, the issue alone resolves 14 and the description 11. A paired sign test is dominated by ties, and where the two differ the description loses more often than it wins. The description's only reliable effect is that its presence in the prompt occasionally pushes the agent to rewrite a file it would otherwise have edited minimally, breaking tasks the issue alone resolves.

\paragraph{A capable model, so the null is not weakness.} A natural objection is that a low-capability agent cannot exploit documentation. We repeat the comparison with Gemini 3.1 Pro on the django Kind-B fixtures, and the null persists: the issue alone resolves 9 of 15, the compact description 8, and the full-length optimized description only 4 (Table~\ref{tab:present}). The longer the documentation, the worse the agent does, because a complete description of a file the agent can already read invites it to rewrite rather than patch; the occasional file-breaking rewrite is the distractor effect that irrelevant context is known to produce \citep{distracted2023}.

\paragraph{An external benchmark, and a positive control.} To rule out our own harness or data as the cause, we move to SWE-ContextBench \citep{swecontextbench2026}, whose Lite split provides 99 related tasks across twelve repositories, with prebuilt per-task Docker environments and an official evaluation harness. We first confirm the instrument is not blind: on the django subset with Gemini 3.1 Pro, the benchmark's own retrieved context resolves 15 tasks to the issue-only condition's 14 on the 29 tasks scored under both, reproducing in direction the effect its authors report. In a separate run on the same django tasks, our compact description resolves 16 tasks to issue-only's 15 on the 32 tasks scored under both. A task is kept only when both of its conditions produce a scorable edit, so the two runs pair slightly different task sets, which is why the issue-only counts differ by one. The compact advantage is a single task, which a paired test cannot distinguish from zero. The harness can see an effect; documentation does not produce one.

\paragraph{The full benchmark at scale.} We then run all three conditions head to head on the full Lite set with Gemini 3.8 Flash under the official Docker evaluation. On the 58 tasks that produced a scorable edit under every condition, the issue alone resolves 33, retrieved context 30, and the compact description 29. The differences are not significant (McNemar's exact test, $p=0.22$ for issue versus compact and $p=0.25$ for issue versus context), and the ordering is consistently against documentation. Only one task is resolved by the compact condition and missed by issue-only, against five in the reverse direction, so documentation does not even rescue a hard subset. The documentation conditions also fail to produce an edit slightly more often. Across five settings, two model families, and three kinds of documentation, the source-present null is uniform. Two of the twelve repositories contribute no scorable task to this run; Appendix~\ref{app:scb} gives the per-repository coverage. This is consistent with a growing body of evidence that added context helps only when it supplies something the model lacks: distilling context to a minimal sufficient subset improves resolution while cutting tokens \citep{ocd2026}, whereas appending text the model does not need can act as a distractor rather than a signal \citep{distracted2023}.

\subsection{Editing without loading the file}

Finally we test whether a description lets the agent edit a file it never loads: the agent names, from the issue and optionally the description, the symbols it needs, receives only those symbols' source, and returns replacements that are spliced in and tested. The result is negative for single files. Attempts with and without the optimized description resolve equally well (17 versus 14 of 33 runs, mean pass fraction 0.94 in both), and the description costs more prompt tokens than it saves, because at single-file scale a faithful description is comparable in size to the file itself. What saves tokens is the targeted protocol, not the description: the issue alone locates the right symbols on all but the smallest files. The one exception marks the boundary. On lambdify, the largest fixture where the description still fits, the agent without it stalls one test short on every attempt while the agent with it resolves all three at fewer tokens than loading the file. Editing steered by a description pays exactly when the code is much larger than its description, which is rare for a single file and is the normal situation at repository scale.

\section{Discussion}

The parts of this paper form one argument, and the experiments settle it in a direction we did not expect at the outset. Documentation helps an agent when it adds what the agent cannot otherwise see. The roundtrip benchmark measures completeness directly and rewards compactness, and it can drive automatic improvement of the description writer: the optimized prompt reaches full regeneration fidelity, and its descriptions carry a module well enough that an agent resolves real issues in it with the source withheld, lifting the mean test-pass fraction from 0.08 to 0.71. But the assumption that motivated the work, that such documentation would also help in the normal setting where the agent can read the code, did not survive a large-scale test. With the source present, neither our static compact descriptions nor the benchmark's own retrieved context beats the issue alone, across a weak local model and a capable hosted one, across our fixtures and an external benchmark, across ten repositories, and under a harness whose sensitivity we verified with a positive control.

The two findings are consistent rather than contradictory. A description that lets us rebuild a file is a restatement of that file, and a restatement adds nothing when the original is already in context; it becomes valuable exactly when the original does not fit. What the experiments add is a sharper boundary: documentation of this kind is a compression format for code that does not fit in the context window, and inside the window it is redundant, occasionally harmful because a full description of a readable file nudges the agent to rewrite rather than patch. This boundary matches independent findings that the value of injected documentation is conditional on the task: domain specifications raise resolution only on specification-dependent bugs and not on generic ones \citep{swebench5g2026}, and correctly selected experience helps while unfiltered context does not \citep{swecontextbench2026}. Our fixtures are standard, well-specified issues whose needed information is already in the issue and the source, so a file-level description is redundant by construction. Length itself is a weaker explanation: although long inputs are known to degrade reliability \citep{contextrot2025}, in our data description length does not predict the penalty, so we treat distraction from redundant content, rather than length alone, as the operative mechanism.

\paragraph{Limitations.} The roundtrip fixture set is small, eleven fixtures plus three hand-built codebases, drawn from a single language and project family, so the benchmark-side numbers characterize the method rather than provide powered measurements. The downstream null, in contrast, is measured across models and repositories, but carries its own caveats: the largest run uses a single economical model and one summary length near eighty words, with the stronger-model replications confined to django subsets; tasks on which some condition produced no edit drop from the paired comparison, and the documentation conditions no-edit slightly more often, so attrition is not perfectly random. The regeneration benchmark is itself model-sensitive: two open-weight models we tried score zero across all fixtures, failing at regeneration by hallucinating imports rather than for any reason tied to description quality, so the benchmark presently requires a sufficiently capable regenerator. The optimization loop is a proof of concept with few fixtures, few iterations, and a single proposer. Our contamination controls reduce but cannot fully eliminate the possibility that training exposure inflates some scores.

\paragraph{Conclusion.} We set out to build documentation that helps coding agents and to test whether it does. The construction succeeded and the hypothesis failed. The roundtrip benchmark treats a description as durable software source, measured by whether code regenerated from it passes the original tests; completeness, not length, drives that fidelity, and a search loop scored by the benchmark discovers a describe prompt that reaches full fidelity, generalizes to held-out files, and rediscovers by optimization the fixes we had found by hand. Those descriptions substitute for source the agent cannot read. They do not help when the agent can read the code, and we report that null at scale rather than around it. The honest summary is a boundary, not a method: documentation of this kind pays when the source cannot be loaded and is redundant when it can. Establishing the same boundary at repository scale, where whole codebases do not fit and the source-withheld regime becomes the default, is the natural next step.

\section*{Acknowledgements}

This work was supported by the National Science Foundation NRT-AI 2244574 and through allocation number CIS240027 from the Advanced Cyberinfrastructure Coordination Ecosystem: Services \& Support (ACCESS) program, which is supported by National Science Foundation grants \#2138259, \#2138286, \#2138307, \#2137603, and \#2138296. The technical support and advanced computing resources from University of Hawaii Information Technology Services -- Research Cyberinfrastructure, funded in part by the National Science Foundation CC* awards \#2201428 and \#2232862 are gratefully acknowledged.

This work was supported by computational resources provided by NPC Labs through B3IQ infrastructure platform.
\bibliography{references}

@article{rtc2024,
  title={Unsupervised Evaluation of Code LLMs with Round-Trip Correctness},
  author={Allamanis, Miltiadis and Panthaplackel, Sheena and Yin, Pengcheng},
  journal={arXiv preprint arXiv:2402.08699}, year={2024}}

@inproceedings{identitychain2024,
  title={Beyond Accuracy: Evaluating Self-Consistency of Code LLMs},
  author={Min, Marcus J. and others},
  booktitle={ICLR}, year={2024}, note={arXiv:2310.14053}}

@inproceedings{swebench2024,
  title={SWE-bench: Can Language Models Resolve Real-World GitHub Issues?},
  author={Jimenez, Carlos E. and others},
  booktitle={ICLR}, year={2024}, note={arXiv:2310.06770}}

@article{rtce2026,
  title={Can LLMs Compress (and Decompress)? Evaluating Code Understanding and Execution via Invertibility},
  author={Maveli, Nickil and Vergari, Antonio and Cohen, Shay B.},
  journal={arXiv preprint arXiv:2601.13398}, year={2026}}

@article{mrgbench2025,
  title={MRG-Bench: Evaluating and Exploring the Requirements of Context for Repository-Level Code Generation},
  author={Li, Haiyang and others},
  journal={arXiv preprint arXiv:2508.02998}, year={2025}}

@inproceedings{codereval2024,
  title={CoderEval: A Benchmark of Pragmatic Code Generation},
  author={Yu, Hao and others},
  booktitle={ICSE}, year={2024}, note={arXiv:2302.00288}}

@article{nl2repobench,
  title={NL2Repo-Bench: Towards Long-Horizon Repository Generation Evaluation of Coding Agents},
  author={Ding, Jingzhe and Long, Shengda and Pu, Changxin and Zhou, Huan and Gao, Hongwan and Gao, Xiang and He, Chao and Hou, Yue and Hu, Fei and others},
  journal={arXiv preprint arXiv:2512.12730}, year={2025}}

@inproceedings{repogenesis,
  title={RepoGenesis: Benchmarking End-to-End Microservice Generation from Readme to Repository},
  author={Peng, Zhiyuan and Yin, Xin and Zhao, Pu and Yang, Fangkai and Wang, Lu and Jia, Ran and Chen, Xu and Lin, Qingwei and Rajmohan, Saravan and Zhang, Dongmei},
  booktitle={Proceedings of the 64th Annual Meeting of the Association for Computational Linguistics (ACL)},
  year={2026}, note={arXiv:2601.13943}}

@article{lessismore2024,
  title={Less is More: DocString Compression in Code Generation},
  author={Yang, Guang and Zhou, Yu and Cheng, Wei and Zhang, Xiangyu and Chen, Xiang and Zhuo, Terry Yue and Liu, Ke and Zhou, Xin and Lo, David and Chen, Taolue},
  journal={arXiv preprint arXiv:2410.22793}, year={2024}}

@article{signatureenough2024,
  title={Do Code Summarization Models Process Too Much Information? Function Signature May Be All That Is Needed},
  author={Ding, Xi and Peng, Rui and Chen, Xiangping and Huang, Yuan and Bian, Jing and Zheng, Zibin},
  journal={ACM Transactions on Software Engineering and Methodology (TOSEM)}, volume={33}, number={6}, year={2024}}

@inproceedings{docreform2024,
  title={Can docstring reformulation with an LLM improve code generation?},
  author={Dainese, Nicola and Ilin, Alexander and Marttinen, Pekka},
  booktitle={Proceedings of the 18th Conference of the European Chapter of the ACL: Student Research Workshop (EACL)},
  pages={296--312}, year={2024}}

@inproceedings{llmlingua2023,
  title={LLMLingua: Compressing Prompts for Accelerated Inference of Large Language Models},
  author={Jiang, Huiqiang and Wu, Qianhui and Lin, Chin-Yew and Yang, Yuqing and Qiu, Lili},
  booktitle={Proceedings of the 2023 Conference on Empirical Methods in Natural Language Processing (EMNLP)},
  year={2023}, note={arXiv:2310.05736}}

@inproceedings{ape2023,
  title={Large Language Models Are Human-Level Prompt Engineers},
  author={Zhou, Yongchao and Muresanu, Andrei Ioan and Han, Ziwen and Paster, Keiran and Pitis, Silviu and Chan, Harris and Ba, Jimmy},
  booktitle={International Conference on Learning Representations (ICLR)}, year={2023},
  note={arXiv:2211.01910}}

@inproceedings{opro2023,
  title={Large Language Models as Optimizers},
  author={Yang, Chengrun and Wang, Xuezhi and Lu, Yifeng and Liu, Hanxiao and Le, Quoc V. and Zhou, Denny and Chen, Xinyun},
  booktitle={International Conference on Learning Representations (ICLR)}, year={2024},
  note={arXiv:2309.03409}}

@inproceedings{dspy2023,
  title={DSPy: Compiling Declarative Language Model Calls into Self-Improving Pipelines},
  author={Khattab, Omar and Singhvi, Arnav and Maheshwari, Paridhi and Zhang, Zhiyuan and Santhanam, Keshav and Vardhamanan, Sri and Haq, Saiful and Sharma, Ashutosh and Joshi, Thomas T. and Moazam, Hanna and Miller, Heather and Zaharia, Matei and Potts, Christopher},
  booktitle={International Conference on Learning Representations (ICLR)}, year={2024},
  note={arXiv:2310.03714}}

@article{promptbreeder2023,
  title={PromptBreeder: Self-Referential Self-Improvement via Prompt Evolution},
  author={Fernando, Chrisantha and Banarse, Dylan and Michalewski, Henryk and Osindero, Simon and Rockt{\"a}schel, Tim},
  journal={arXiv preprint arXiv:2309.16797}, year={2023}}

@article{textgrad2024,
  title={TextGrad: Automatic ``Differentiation'' via Text},
  author={Yuksekgonul, Mert and Bianchi, Federico and Boen, Joseph and Liu, Sheng and Huang, Zhi and Guestrin, Carlos and Zou, James},
  journal={arXiv preprint arXiv:2406.07496}, year={2024}}

@inproceedings{docgpt3,
  title={Automatic Code Documentation Generation Using GPT-3},
  author={Khan, Junaed Younus and Uddin, Gias},
  booktitle={Proceedings of the 37th IEEE/ACM International Conference on Automated Software Engineering (ASE)},
  year={2022}, note={arXiv:2209.02235}}

@article{chatgptsumm,
  title={Automatic Code Summarization via ChatGPT: How Far Are We?},
  author={Sun, Weisong and Fang, Chunrong and You, Yudu and Miao, Yun and Liu, Yi and Li, Yuekang and Deng, Gelei and Huang, Shenghan and Chen, Yuchen and Zhang, Quanjun and others},
  journal={arXiv preprint arXiv:2305.12865}, year={2023}}

@inproceedings{doccompare,
  title={A Comparative Analysis of Large Language Models for Code Documentation Generation},
  author={Dvivedi, Shubhang Shekhar and Vijay, Vyshnav and Pujari, Sai Leela Rahul and Lodh, Shoumik and Kumar, Dhruv},
  booktitle={Proceedings of the 1st ACM International Conference on AI-Powered Software (AIware)},
  pages={65--73}, year={2024}, note={arXiv:2312.10349}}

@misc{autoagent,
  title={autoagent}, author={Gu, Kevin R.}, howpublished={\url{https://github.com/kevinrgu/autoagent}}, year={2024}}

@misc{metaharness,
  title={metaharness}, author={{SuperagenticAI}}, howpublished={\url{https://github.com/SuperagenticAI/metaharness}}, year={2024}}

@article{humaneval2021,
  title={Evaluating Large Language Models Trained on Code},
  author={Chen, Mark and Tworek, Jerry and Jun, Heewoo and Yuan, Qiming and others},
  journal={arXiv preprint arXiv:2107.03374}, year={2021}}

@article{mbpp2021,
  title={Program Synthesis with Large Language Models},
  author={Austin, Jacob and Odena, Augustus and Nye, Maxwell and Bosma, Maarten and others},
  journal={arXiv preprint arXiv:2108.07732}, year={2021}}

@article{lostmiddle2023,
  title={Lost in the Middle: How Language Models Use Long Contexts},
  author={Liu, Nelson F. and Lin, Kevin and Hewitt, John and Paranjape, Ashwin and Bevilacqua, Michele and Petroni, Fabio and Liang, Percy},
  journal={Transactions of the Association for Computational Linguistics},
  year={2023},
  note={arXiv:2307.03172}}

@inproceedings{sweagent2024,
  title={SWE-agent: Agent-Computer Interfaces Enable Automated Software Engineering},
  author={Yang, John and Jimenez, Carlos E. and Wettig, Alexander and Lieret, Kilian and Yao, Shunyu and Narasimhan, Karthik and Press, Ofir},
  booktitle={Advances in Neural Information Processing Systems (NeurIPS)},
  year={2024}, note={arXiv:2405.15793}}

@inproceedings{autocoderover2024,
  title={AutoCodeRover: Autonomous Program Improvement},
  author={Zhang, Yuntong and Ruan, Haifeng and Fan, Zhiyu and Roychoudhury, Abhik},
  booktitle={Proceedings of the 33rd ACM SIGSOFT International Symposium on Software Testing and Analysis (ISSTA)},
  year={2024}}

@misc{openwiki2026,
  title={OpenWiki: An Open-Source Agent for Repository Documentation},
  author={{LangChain}},
  year={2026},
  howpublished={\url{https://github.com/langchain-ai/openwiki}},
  note={Software; announced at \url{https://www.langchain.com/blog/introducing-openwiki-an-open-source-agent-for-repo-documentation}}}

@article{swecontextbench2026,
  title={SWE Context Bench: A Benchmark for Context Learning in Coding},
  author={Zhu, Jiayuan and others},
  journal={arXiv preprint arXiv:2602.08316}, year={2026}}

@article{contextbench2026,
  title={ContextBench: A Benchmark for Context Retrieval in Coding Agents},
  author={Li, Han and Zhu, Letian and Zhang, Bohan and Feng, Rili and Wang, Jiaming and Pan, Yue and Barr, Earl T. and Sarro, Federica and Chu, Zhaoyang and Ye, He},
  journal={arXiv preprint arXiv:2602.05892}, year={2026}}

@article{swebench5g2026,
  title={SWE-Bench 5G: Benchmarking AI Coding Agents on Telecom Network Engineering Tasks},
  author={Chen, Jiao and Tang, Jianhua and Yang, Xiaotong and Lv, Zuohong},
  journal={arXiv preprint arXiv:2604.26278}, year={2026}}

@article{ocd2026,
  title={Compressing Code Context for LLM-based Issue Resolution},
  author={Jia, Haoxiang and Barr, Earl T. and Mechtaev, Sergey},
  journal={arXiv preprint arXiv:2603.28119}, year={2026}}

@inproceedings{distracted2023,
  title={Large Language Models Can Be Easily Distracted by Irrelevant Context},
  author={Shi, Freda and Chen, Xinyun and Misra, Kanishka and Scales, Nathan and Dohan, David and Chi, Ed and Sch\"arli, Nathanael and Zhou, Denny},
  booktitle={International Conference on Machine Learning (ICML)}, year={2023}, note={arXiv:2302.00093}}

@techreport{contextrot2025,
  title={Context Rot: How Increasing Input Tokens Impacts LLM Performance},
  author={Hong, Kelly and Troynikov, Anton and Huber, Jeff},
  institution={Chroma}, year={2025}, note={Technical Report}}
\newpage
\appendix
\appendix
\section{Testing Whether Length Hurts}
\label{app:contextrot}

The main text treats description length as second-order: what determines an agent's success is completeness, not length. This appendix reports the experiments behind that conclusion, including a case where an apparent length effect turned out to be an artifact of our setup.

\paragraph{Setup.} All experiments here reuse the comprehension task from the main text, in which the needed facts appear only in the description and the implementation is a trivial lookup, so success depends only on whether the description conveyed the facts. We compare a compact description, which lists the facts tightly, against a verbose description, which carries the same facts embedded in much longer prose. We measure the fraction of independent attempts that recover all the checked values.

\paragraph{Equal uplift across model scale.} A natural worry is that a long description hurts a weaker model, even when it is complete, because models use information less reliably as the surrounding context grows \citep{lostmiddle2023}. We first tested this with five short facts, comparing a 38-word description against a 664-word one carrying the same content, across models of decreasing size. The result was the same at every size (Table~\ref{tab:contextrot-easy}): no description scored zero, and both the compact and the verbose description scored full marks. Length did not hurt.

\begin{table}[h]
\centering
\begin{tabular}{@{}lccc@{}}
\toprule
Agent model & No desc. & Compact (38w) & Verbose (664w) \\
\midrule
Flash-lite   & 0/8 & 8/8 & 8/8 \\
Gemma 4 31B  & 0/8 & 8/8 & 8/8 \\
Gemma 3 4B   & 0/8 & 8/8 & 8/8 \\
\bottomrule
\end{tabular}
\caption{Five-fact task. A compact and a verbose description carrying the same facts give identical uplift, from Flash-lite down to a 4B model.}
\label{tab:contextrot-easy}
\end{table}

\paragraph{A harder task, and a confound.} We then made the task harder to give any length effect a better chance to appear. We used 28 settings instead of five, including deliberately confusable key families (for example \texttt{max\_retries}, \texttt{max\_retry\_delay}, \texttt{connection\_max\_retries}, and \texttt{request\_max\_retries}), and checked ten specific values that sit among their confusable neighbours. Our first verbose description described each key in words but never wrote the exact key string, while the compact version listed the keys verbatim. On this version the verbose description collapsed to zero while compact stayed perfect, which looked like a strong effect of length.

The comparison was confounded, however. The verbose description did not actually contain the exact identifiers the task required, so its failure could simply be missing information rather than degradation over a long context. To separate the two, we wrote a fair verbose description that contains the exact key strings, still embedded in long filler with the confusable keys scattered across paragraphs. With that version the effect disappears (Table~\ref{tab:contextrot-hard}): the fair verbose description passes as well as the compact one, on every model we tried, and the models correctly distinguish the confusable keys.

\begin{table}[h]
\centering
\small
\begin{tabular}{@{}lccc@{}}
\toprule
Agent model & Compact & Verbose & Verbose \\
            & (listed) & (described) & (buried) \\
\midrule
Flash-lite         & 8/8 & 0/8 & 8/8 \\
Gemini 2.5 Flash   & 4/4 & 0/4 & 4/4 \\
Gemma 3 4B         & 8/8 & 0/8 & 8/8 \\
\bottomrule
\end{tabular}
\caption{Harder task with a confound control. The verbose description fails only when it omits the exact key strings the task needs. Once those strings are present, even buried in long filler, it passes as well as the compact one.}
\label{tab:contextrot-hard}
\end{table}

\paragraph{Reading.} The apparent length effect was a completeness effect in disguise. A verbose description fails when it leaves out something the task needs, not because it is long. When the verbose description is complete, it matches the compact one, even for a small model and even amid heavy filler and distractors. This is why the main text treats length as second-order and completeness as the property that decides an agent's success.

\section{Fixture Selection Criteria}
\label{app:filter}

The benchmark fixtures are chosen by an explicit, scripted filter rather than by hand, so the selection is reproducible and free of per-instance judgement. We record here each criterion and its motivation.

\begin{itemize}
\item \textbf{Single non-test source file in the gold patch.} An instance qualifies only if its gold patch modifies exactly one non-test source file. This keeps the roundtrip target well defined: the describe and regenerate stages act on one module, and the oracle attributes a pass or fail to that module rather than to an interaction across several edited files.
\item \textbf{Oracle tests named by the test patch.} The test patch must name the tests that certify the fix. These become the oracle for both the roundtrip and the resolution experiments, so an instance without an identifiable oracle cannot be scored and is excluded.
\item \textbf{Subpackage size bound (three to fifteen files).} For the multi-file setting a subpackage must contain between three and fifteen source files. The lower bound removes trivial packages with no cross-file structure worth describing; the upper bound keeps a package small enough that its regeneration and evaluation fit a single run and context window.
\item \textbf{Self-contained tests.} A subpackage must own its tests, either in an inner \texttt{tests/} directory or in name-matched test files, so the oracle can run against the subpackage in isolation without pulling in unrelated suites.
\item \textbf{Directory exclusions.} Test directories, private directories, vendored code, and installed \texttt{site-packages} are excluded from the source count and from selection, since they are not the project's own maintained source and would distort both the density measure and the regeneration target.
\item \textbf{Executable baseline.} We build an environment for each candidate, apply the gold and test patches, and keep the instance only if its original code passes its own oracle in full. Instances whose baseline fails to build or pass, or whose era predates a current Python toolchain, are excluded and logged with the reason.
\end{itemize}

Applying the single-file and named-oracle criteria admits 429 of the 2294 SWE-bench Verified instances; the remaining project, era, and executability constraints reduce the set to the fixtures used in the main text. Each threshold is a one-line rule in the selection script, so none of the numbers reflects a discretionary choice.

\section{Source-Present Evaluation: Reproducibility and Per-Repository Results}
\label{app:scb}

\paragraph{Protocol.} The source-present experiments on SWE-ContextBench Lite use the benchmark authors' prebuilt per-task Docker images, one per instance, each containing the repository checked out at the instance's pre-fix commit together with the era-correct interpreter and dependencies. For each task we pull the image, extract the repository, restore the target file to its pre-fix state, and run the agent in one of three conditions: the issue alone; the issue with the benchmark's retrieved past-task context; or the issue with a static compact description of the target file. The compact description is generated from the pre-fix file with the optimized describe prompt (Appendix~\ref{app:prompt}) and compressed to roughly eighty words, so it depends only on the file and cannot leak the fix. The agent edits in place; we take the resulting diff as the predicted patch and score it with the benchmark's official evaluation harness, which applies the test patch and requires every fail-to-pass and pass-to-pass test to pass inside the container. A task is counted only when it produces a scorable edit under every condition being compared. Scoring the gold patch through the same harness recovers full resolution, which confirms that the pipeline is faithful. The code, task list, generated descriptions, and per-task results are released with the paper.

\paragraph{Per-repository issue-only results.} Table~\ref{tab:perrepo} breaks the full issue-only run down by repository. Resolution varies widely across projects, from matplotlib at eight of nine to sympy at six of twenty-one, which reflects task difficulty and era rather than anything about documentation. The Lite split spans twelve repositories, but two contribute no scorable task: the single pylint-dev task produced no edit, and both pytest-dev tasks failed in the evaluation harness, so ten repositories and 68 of the 99 tasks appear in the table; the head-to-head comparisons in the main text are computed within the same task set, so this variation is controlled for.

\begin{table}[t]
\centering
\begin{tabular}{@{}lc@{}}
\toprule
Repository & Resolved / Scored \\
\midrule
matplotlib   & 8/9 \\
django       & 14/22 \\
scikit-learn & 3/5 \\
psf (requests) & 2/3 \\
sympy        & 6/21 \\
astropy      & 1/2 \\
mwaskom (seaborn) & 1/2 \\
pydata (xarray) & 1/2 \\
sphinx-doc   & 1/1 \\
pallets (flask) & 0/1 \\
\midrule
Total        & 37/68 \\
\bottomrule
\end{tabular}
\caption{Issue-only resolution on SWE-ContextBench Lite by repository (Gemini 3.8 Flash, official Docker evaluation), over the 68 of 99 tasks that produced a scorable edit. pylint-dev (one task) and pytest-dev (two tasks) produced no scorable edit and are omitted.}
\label{tab:perrepo}
\end{table}

\section{The Discovered Describe Prompt}
\label{app:prompt}

The optimization section of the main paper reports that the optimization loop discovers a describe prompt that reaches full fidelity and independently rediscovers the fixes for the failure modes catalogued there. For reference and reproducibility, the discovered prompt is reproduced below verbatim.

\begin{scriptsize}
\begin{verbatim}
You are an expert programmer creating a high-fidelity,
compact, natural-language specification of a Python file.
Your output is a blueprint for a code-generation AI, which
must be able to losslessly recreate the original file and
pass all its tests from your description alone.

Structure your description as follows:

1. Module-Level Preamble:
   - Imports: List all imports precisely, e.g.,
     `from module import name as alias`.
   - Constants & Globals: Define all module-level constants,
     variables, and type aliases with their exact names and
     literal values. For complex data structures (dicts,
     lists), detail their full contents.

2. Code Objects (Classes and Functions):
   For each class and function, including private _ members:
   - Header: For a function, state its exact signature
     (name, parameters, type hints, precise default values).
     For a class, state its name, base classes, any metaclass.
   - Attributes (classes only): List all class and instance
     attributes, their types, and how/where initialized.
   - Implementation Logic: Describe the step-by-step logic;
     what the code does, not a line-by-line translation.
     Detail control flow and the conditions governing it.
     Specify key operations: data transformations,
     calculations, algorithms, calls to other functions.
     State the return value(s), their types, and the
     conditions for each return path. Document any exceptions
     raised and the exact conditions that trigger them.

Guiding Principles:
   - Unambiguous & Concise: prioritize technical accuracy
     for perfect functional replication.
   - Omit the Obvious: do not describe standard language
     features or boilerplate.
   - No High-Level Explanations: avoid conceptual overviews,
     design rationale, or code comments.
\end{verbatim}
\end{scriptsize}

The instructions it converged on map directly onto the hand-identified failure modes of the main paper: it asks for all imports to be listed, for module-level constants to be given with their exact literal values, and for exact signatures, defaults, and the conditions of every exception and return path.

\section{Base Prompts}
\label{app:baseprompts}

For completeness we give the base prompts used before optimization. The describe stage used the following system prompt.

\begin{scriptsize}
\begin{verbatim}
You are given the complete source of a small Python
program. Write a precise natural-language specification
of its behavior: every module-level function and class,
including any whose names begin with an underscore, since
callers and tests may import them directly. Cover each
one's inputs, outputs, and edge cases, detailed enough
that another engineer could reimplement it from your
description alone, without seeing the code. Describe
behavior, not a line-by-line transcription. Do not
include or infer test code.
\end{verbatim}
\end{scriptsize}

The regenerate stage used the following system prompt, which sees only the description and the project scaffold.

\begin{scriptsize}
\begin{verbatim}
You are given a natural-language specification and a
project scaffold. Implement the program so it satisfies
the specification. Output every source file you create as
a block in exactly this format:
=== <relative/path> ===
<full file contents>
Start each file with its own '=== path ===' header line.
Follow the scaffold's layout (e.g. packages under src/).
Do not write tests.
\end{verbatim}
\end{scriptsize}

The optimization loop of the main paper uses a third prompt, which drives the proposer. The proposer is shown the current describe prompt together with its measured fidelity and length, and is asked to rewrite it. Its system prompt is the following.

\begin{scriptsize}
\begin{verbatim}
You are improving the system prompt used to DESCRIBE a
code file in natural language, so another model can
regenerate the code from the description alone and pass
the original tests. Goal: maximize regeneration fidelity
while keeping the description compact. Given the current
prompt and its measured fidelity and length, rewrite it
to score higher. Output ONLY the new prompt text, no
preamble.
\end{verbatim}
\end{scriptsize}

The accompanying user message supplies the current prompt and the scores it achieved, in the following template.

\begin{scriptsize}
\begin{verbatim}
Current describe prompt:
<current prompt>

Measured: mean fidelity <fidelity>, mean words <words>.
Per fixture (name, fidelity, words): <per-fixture scores>

Rewrite to improve fidelity while staying compact.
Output only the new prompt.
\end{verbatim}
\end{scriptsize}

The proposer is \textsc{gemini-2.5-pro} sampled at temperature $0.4$, so that successive proposals differ. The describe and regenerate stages scored inside the loop remain at temperature zero, so each candidate prompt receives a deterministic score.

\section{OpenWiki Baseline Details}
\label{app:openwiki}

This appendix gives the setup behind the OpenWiki comparison in the main text. OpenWiki \citep{openwiki2026} is distributed as a command-line tool that generates a documentation set for a repository. It ships with a fixed set of model providers and does not include our benchmark's model. Because it is open source, we pointed its provider configuration at the same model family our benchmark uses, so that any difference in the comparison reflects the documentation rather than the model. We then ran it on each fixture to produce documentation, and fed that documentation into the same regenerate and evaluate stages used for our own descriptions.

For the regeneration comparison, the documentation OpenWiki produced was used in place of the describe stage's output; everything downstream was identical to a normal run. OpenWiki's documentation is organized for a human reader, with an overview, usage examples, and an API summary. On \texttt{saferepr} it described the public functions but not the private helper the tests import, so the regenerated module was missing that symbol and failed at import. On \texttt{unitsystem} it omitted a class attribute the code relies on, with the same effect. On \texttt{contains} it produced runnable code that passed a majority of tests but missed two.

For the agent-task comparison we used the deny-override fixture, in which the policy engine's code trusts the order in which rules are passed and never implements the priority-and-deny-override contract; that contract exists only in our description. OpenWiki, documenting the code as written, reported accurately that the engine ignores priority and evaluates rules in input order. This is the observed behavior but the opposite of the contract the task requires, so an agent given OpenWiki's documentation reproduces the order-dependent behavior and fails the task, exactly as it does with no documentation at all. The comparison therefore isolates the distinction the paper draws throughout: our descriptions supply the contract an agent needs, while documentation of observed behavior does not.

\section{An Example Description}
\label{app:exampledesc}

For concreteness, the describe stage's output for the \texttt{contains} fixture is reproduced below in full. This description regenerates its file at fidelity 1.00; note that it specifies exact argument positions (\texttt{self.args[1]}), the exact error message string, and the filtering logic of a property, the level of internal detail that human-facing documentation typically omits.

\begin{scriptsize}
\begin{verbatim}
# Module: sympy.sets.contains

This module defines the Contains class, which represents
the boolean assertion that an element x is a member of a
set S.

## Class: Contains

Contains inherits from BooleanFunction and represents the
mathematical statement x in S.

### Class Method: eval

@classmethod
def eval(cls, x, s)

Evaluates the membership assertion of an element x in a
set s.
- Inputs: x, any SymPy expression representing the
  element; s, an instance of Set representing the set.
- Returns: S.true or S.false if membership can be
  definitively determined; an instance of Set if the
  evaluation of s.contains(x) results in a Set; None if
  the relation cannot be resolved to a boolean constant
  or a set, which signals SymPy to construct and return
  an unevaluated Contains(x, s) object.
- Raises: TypeError if s is not an instance of Set. The
  error message is "expecting Set, not <type_name>",
  where <type_name> is the class name of s.

### Property: binary_symbols

Returns a Python set of the binary symbols present within
the definition of the set s (the second argument of
Contains).
- Iterates through the arguments of the set s
  (self.args[1].args).
- Filters these arguments to include only those elements
  i where i.is_Boolean is True, i.is_Symbol is True, or i
  is an instance of Eq or Ne.
- Computes the union of the binary_symbols property of
  all such filtered arguments and returns the resulting
  set.

### Method: as_set

def as_set(self)

Returns the set associated with this membership
assertion.
- Returns: the Set instance s (the second argument of the
  Contains expression, self.args[1]).
\end{verbatim}
\end{scriptsize}

\end{document}